\documentclass{article}
\usepackage{spconf,amsmath,graphicx}
\usepackage{makecell}
\usepackage[table]{xcolor}
\usepackage{color}

\newcommand{\convMatrix}[3]{
\begin{math}
    \begin{bmatrix}
        \text{conv},\ 3 \times 3,\ #1 \\
        \text{conv},\ 3 \times 3,\ #2 \\ 
        \text{conv},\ 3 \times 3,\ #3 
    \end{bmatrix}
\end{math}
}

\newcommand{\convRow}[4]{$ \text{#1},\ #2 \times #3,\ #4 $}
\usepackage{multirow}

\title{Detecting Multiple Speech Disfluencies using a Deep Residual Network with Bidirectional Long Short-Term Memory}

\name{Tedd Kourkounakis, Amirhossein Hajavi, Ali Etemad}
\address{Department of Electrical and Computer Engineering\\Queen's University, Kingston, Ontario, Canada\\ 
\small{\texttt{\{tedd.kourkounakis, a.hajavi, ali.etemad\}@queensu.ca}}}

\begin{document}

\maketitle

\begin{abstract}
Stuttering is a speech impediment affecting tens of millions of people on an everyday basis. Even with its commonality, there is minimal data and research on the identification and classification of stuttered speech. This paper tackles the problem of detection and classification of different forms of stutter. As opposed to most existing works that identify stutters with language models, our work proposes a model that relies solely on acoustic features, allowing for identification of several variations of stutter disfluencies without the need for speech recognition. Our model uses a deep residual network and bidirectional long short-term memory layers to classify different types of stutters and achieves an average miss rate of 10.03\%, outperforming the state-of-the-art by almost 27\%. 
\end{abstract}

\section{Introduction}
\label{sec:intro}

Speech disfluencies are inconsistencies and interruptions in the flow of otherwise normal speech. Of these speech impediments, stuttering is one of the most prominent, affecting over 70 million people, about one percent of the global population \cite{stutteringfoundation}. 5-10\% of children stutter at some point in their childhood, with a quarter of these children maintaining their stutters throughout their entire lives \cite{nidcd}. Common therapy methods often involve helping the patient monitor and maintain awareness of their speaking patterns in order to correct them \cite{stutteringbook}. Moreover, therapeutic success rates have been reported to be over 80\%, especially when detected and dealt with in early stages \cite{Saltuklaroglu2004}. 
Accordingly, with the recent advances in machine learning, deep learning, and language/speech processing techniques, developing smart and interactive tools for detection and therapy is now a real possibility.

In addition to interactive therapy purposes, other applications can be realized for automated stutter recognition. Fluent speech is crucial and influential in presentations such as talks and business communications \cite{morreale2000}. There are currently a number of applications available to assist speakers in monitoring and improving their presentation skills. For example, monitoring of features like volume, rate of speech, and intonation, among others have been explored in this context \cite{ROBOCOP} \cite{Kotz2003}. However, detection and quantification of stutters has not yet been fully explored for such applications.

Despite the many potential applications for automated stutter detection, little research has been done in this area. This is partially due to the fact that the notion of detecting and classifying the type and location of stutters can be a difficult problem, especially when factoring in variables such as gender, speech rate, accent, and phone-realization \cite{speechfactors}. Existing works in the area mostly rely on automatic speech recognition (ASR) to first convert audio signals to text, and then utilize language models to detect and identify the stutters  \cite{Heeman2016} \cite{interspeech2017} \cite{interspeech2018}. While this approach has proven effective and achieved promising results, the reliance on ASR can both be a potential source for error, as well as an unnecessary additional computational step.

\begin{figure}[t]
    \begin{center}
    \includegraphics[width=0.8\columnwidth]{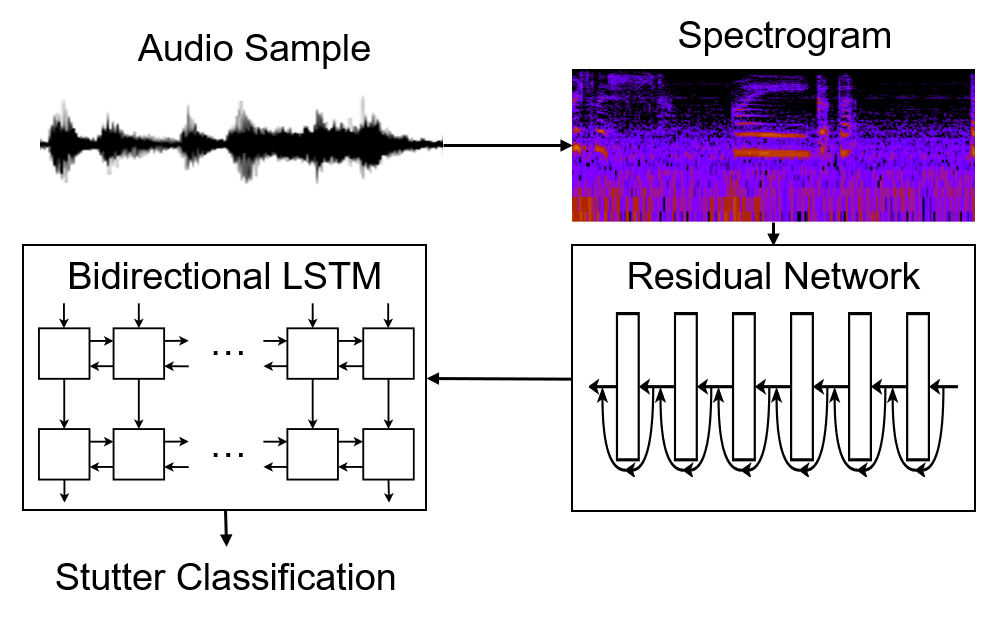}
    \end{center}
\caption{Proposed stutter detection system diagram.}
\label{fig:intro_diagram}
\end{figure}

In this paper, we propose a model that directly utilizes audio speech signals to detect and classify stutters, skipping the ASR step and the need for language models. Our method uses spectrogram features to train a deep neural network with residual layers followed by bidirectional long short-term memory (Bi-LSTM) units to learn to locate and identify different types of stutters. The overview of our method is presented in Figure \ref{fig:intro_diagram}. Our experiments show the effectiveness of our approach in generalizing across multiple classes of stutters while maintaining a high accuracy and strong consistency between classes.

\section{Related Work}
\label{sec:format}
Early studies on the topic focused on the feasibility of stutter differentiation, with training and testing often being performed on a small set of specific stuttered words. For example, a hidden Markov model (HMM) was used to create a stutter recognition assistance tool \cite{tan2007}. Testing results averaged to 96\% and 90\% accuracy on human and artificially generated stuttered speech samples respectively for a single pre-determined word \cite{tan2007}.

A number of assumptions are often made in order to simplify the problem of disfluency detection. For example, as different disfluencies vary heavily by nature, proposed solutions often tackle one single type of stutter (such as interjections, prolongations, or repetitions) at a time. In \cite{ravikumar2009}, for instance, sound repetition stutters were accurately detected on a small set of trained words. Another common assumption used for simplification has been to remove under-represented subject classes (for example based on gender or age) \cite{ravikumar2009}, \cite{Chee2009}, \cite{interspeech2018}.

As ASR and natural language processing (NLP) has evolved greatly in recent years, such methods have become increasing popular for the problem of stutter classification and recognition. One such method incorporated annotations from speech language pathologist to a word lattice model, improving the baseline method by a relative 7.5\% \cite{Heeman2016}. Another model using Bi-LSTMs with condition random fields (CRFs) to get an average F-score of 85.9\% across all stutter types \cite{zayats2016}. The current state-of-the-art stutter classification method uses task-oriented finite state transducer (FST) lattices to detect repetition stutters with an average 37\% miss rate across 4 different types of \cite{interspeech2018}.

\section{Proposed Network}
Our proposed method first generates spectrogram feature vectors from the audio clips. The spectrograms are then passed through a deep residual neural network, mapping the spectrogram matrices to a linear vector. These are then be passed through a bidirectional LSTM to learn the extracted feature embeddings for different types of stutters. Following the different steps of our proposed pipeline are described.

\subsection{Feature extraction}
Spectrograms are commonly used features in speech analysis in different applications ranging from speech recognition to noise cancellation \cite{Kingsbury1998} \cite{Szabelska2013}. We use spectrograms as the sole feature for our model. These features are generated every 10 \textit{ms} on a 25 \textit{ms} window for each 4-second audio clip.

\subsection{Feature embedding layers} 
\begin{figure*}[!t]
    \begin{center}
    \includegraphics[width=0.9\linewidth]{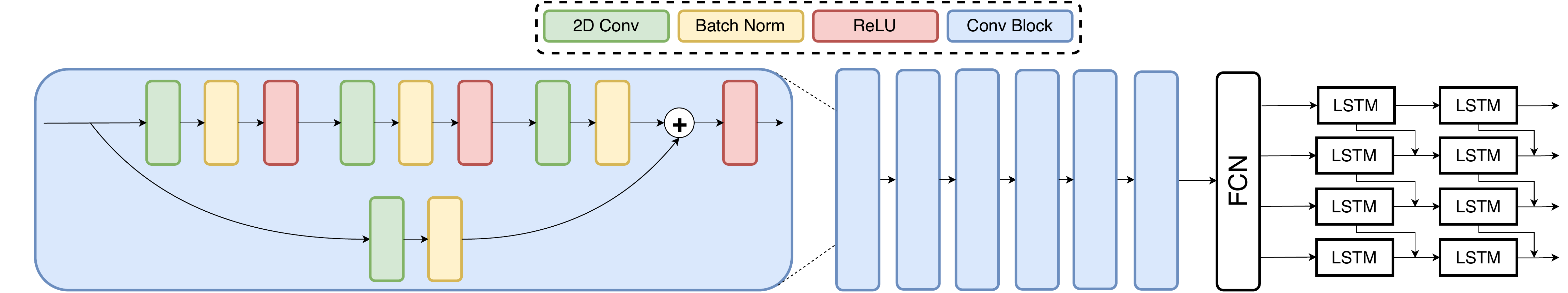}
    \end{center}
\caption{The architecture of our model.}
\label{fig:res}
\end{figure*}

We utilize a residual network \cite{resnet} in our model in order to effectively learn the stutter-specific features while avoiding issues such as the vanishing gradient problem. The use of this type of network also allows a deep architecture (a depth of 18 convolution layers) without overfitting, especially when considering the relatively small size of the dataset. Moreover, architectures with residual components have recently shown considerable promise in speech analysis \cite{xiong2018microsoft} \cite{hajavi2019}. In our proposed solution, each group of 3 convolutional layers is referred to as a convolutional block. Figure \ref{fig:res} presents the convolutional blocks and the stacked blocks in our mode. We used batch normalization and ReLu activation functions in the model. Table \ref{table:hyperparameters} presents the hyperparameters of our network. The detection task for each stutter is formulated as a binary problem, with the same architecture mentioned being used for every disfluency type.

\subsection{Recurrent layers}
The learned feature embeddings are provided to 2 recurrent layers, each consisting of 512 bidirectional LSTM units \cite{BiLSTM}. We utilized LSTM layers as they have been proven to be effective in classification when dealing with short sequential data, and are a popular approach in speech and NLP \cite{LSTM}. In the context of the problem at hand, most stutters tend to be quick and last only a fraction of the 4-second audio clip that they are contained in. Therefore the LSTM layers don't suffer from memory issues \cite{dropout}. Lastly, the use of \textit{bidirectional} LSTMs allow the model to learn both past and future embeddings, providing further context for our problem. Dropout rates of 0.2 and 0.4 are utilized after each recurrent layer.

\section{Experiment Setup and Results}
\label{sec:pagestyle}
\subsection{Data and annotation}
Speech samples were collected from the University College London’s Archive of Stuttered Speech (UCLASS) Release One \cite{UCLASS} dataset, created by the Division of Psychology and Language Sciences within the university. The dataset contains samples of monologues from 139 participants, ranging between 8 and 18 years of age, with known stuttered speech impediments of different severity. Of these recordings, 25 unique participants were used due to the availability of their orthographic transcriptions of the monologues.

\begin{table}
\caption{Our ResNet architecture and hyper-parameters.}
\label{table:hyperparameters}
\resizebox{\columnwidth}{!}
{
\begin{tabular}{c|c|c}
\hline
Module      & Input Spectrogram $(257 \times T \times 1)$ & Output Size \\[.1cm] \hline
\multirow{16}{*}{Conv Module}
            & \convRow{Conv2D}{7}{7}{64} & $ 257 \times T \times 64 $ \\[.1cm]
\cline{2-3} & \convMatrix{32}{64}{64} & $ 128 \times T/2  \times  64 $ \\[.1cm]
\cline{2-3} & \convMatrix{64}{128}{128} & $  64 \times T/4  \times 128 $ \\[.1cm]
\cline{2-3} & \convMatrix{128}{128}{128} & $  64 \times T/8  \times 128 $ \\[.1cm]
\cline{2-3} & \convMatrix{128}{64}{64} & $  32 \times T/16 \times 64 $ \\[.1cm]
\cline{2-3} & \convMatrix{64}{64}{32} & $  16 \times T/32  \times 32 $ \\[.1cm]
\cline{2-3} & \convMatrix{32}{16}{16} & $  8 \times T/64 \times 16 $ \\[.1cm]
\hline
\end{tabular}
}
\end{table}

Forced time-alignment was used on the audio and transcriptions to generate a timestamp for each word and stutter spoken \cite{timealign}. The stutter annotation approach is similar to previously used methods \cite{yairiambrose, justeandrade}. We then manually annotated each recording for one of 7 stutter disfluencies \cite{Yaruss1997}: sound repetition, word repetition, phrase repetition, revision, interjection, or prolongation. A description of each type of stutter can be found in Table \ref{table: stutters}. We leave out part-word repetition disfluencies as the dataset contained only few samples of such stutters, preventing our deep learning approach from properly learning the classification task. Each monologue recording was segmented into 4-second samples, totalling to 800 labeled audio clips.

\begin{table*}
\begin{center}
\small
\caption{Types of stutter considered.}
\label{table: stutters}
\begin{tabular}{l l l l} 
\hline
Label & Stutter Disfluency & Description & Example \\
\hline\hline

S & Sound Repetition & Repetion of any phenome & th-th-this \\
\hline
W & Word Repetition & Repetition of any word & why why \\
\hline
PH & Phrase Repetition & Repetition of multiple successive words &  I know I know that\\
\hline
R & Revision & Repetition of thought, rephrased mid sentence & I think that- I believe that \\
\hline
I & Interjection & Fabricated word or sounds, added to stall for time & um, uh\\ 
\hline
PR & Prolongation & Prolonged sounds & whoooooo is it \\
\hline
\end{tabular}
\end{center}
\end{table*}

\subsection{Implementation details}
The model was built using TensorFlow's Keras API \cite{chollet2015keras}. It was trained  with a learning rate of $10^{-4}$ over 30 epochs, with minimal improvement in results seen in following epochs. A root means square propagation (RMSProp) optimizer was used, as well as the softmax loss function. An Nvidia 1080 Ti GPU was used to perform the training.

\begin{table*}
\begin{center}
\footnotesize
\caption{The percentage miss-rate (MR) and accuracy (Acc) for each stutter type is presented using LOSO validation.}
\label{table: results}
\begin{tabular}{cc|cc|cc|cc|cc|cc|cc}
\hline
 & & \multicolumn{2}{c|}{S} & \multicolumn{2}{c|}{W} & \multicolumn{2}{c|}{PH} & \multicolumn{2}{c|}{R} & \multicolumn{2}{c|}{I} & \multicolumn{2}{c}{PR} \\
Paper & Method & MR & Acc & MR & Acc & MR & Acc & MR & Acc & MR & Acc & MR & Acc\\ \hline\hline
Alharbi et al. \cite{interspeech2018} & Word Lat. & 60 & -- & \textbf{0} & -- & \cellcolor{gray!25} & \cellcolor{gray!25} & 25 & -- & \cellcolor{gray!25} & \cellcolor{gray!25} & \textbf{0} & -- \\
Ours (baseline) & ResNet+LSTM & 20.13 & 83.20 & 3.40 & 95.60 & 4.93 & 95.07 & 3.00 & 96.99 & 25.31 & 80.80 & 6.12 & 93.88 \\
\textbf{Ours (proposed)} & \textbf{ResNet+Bi-LSTM} & \textbf{18.10} & \textbf{84.10} & 3.20 & 96.60 & \textbf{4.46} & \textbf{95.54} & \textbf{2.86} & \textbf{97.14} & \textbf{25.12} & \textbf{81.40} & 5.92 & 94.08\\\hline
\end{tabular}%
\end{center}
\end{table*}

\begin{table}
\begin{center}
\footnotesize
\caption{Average accuracy and miss-rate of stutter classification models.}
\label{table: results_ave}
\begin{tabular}{c c c c} 
\hline
Paper & Method & Ave. MR & Ave. Acc \\
\hline\hline
Alharbi et al. \cite{interspeech2018} & Word Lat. & 37\% & -- \\
Ours (baseline) & ResNet+LSTM & 10.45\% & 90.96\% \\
\textbf{Ours (proposed)} & \textbf{ResNet+Bi-LSTM} & \textbf{10.03\%} & \textbf{91.15\%} \\
\hline
\end{tabular}
\end{center}
\end{table}

\subsection{Validation}
To rigorously test our proposed model, leave-one-subject-out (LOSO) cross validation was used: the model was trained on the speech of 24 of the UCLASS participants, while the last subject's audio was used for testing. The process was repeated 25 times, testing the model on a different subject every time. For this dataset, accuracy (Acc) and miss rate (MR) values have been reported in prior work. Lastly, it should be noted that we target 6 categories of stutter disfluencies, as opposed to most prior work where fewer classes are considered.

\subsection{Performance and Comparison}
The results of our experiments for the UCLASS dataset is summarized in Table \ref{table: results}, where we compare our method to  \cite{interspeech2018}. Additionally, to evaluate the need for \textit{bidirectional} LSTM as opposed to a unidirectional LSTM, we compare our results to a baseline model where a ResNet with LSTM is used instead of our proposed model. The table shows that our method outperforms the state-of-the-art in detection of sound repetition and revisions by considerable margins (an improvements of 41.90\% and 22.14\% respectively).

The statistical language models and task-oriented word lattices used in other methods rely heavily on generating a strong orthographic transcriptions for each speaker. As a result, while these methods struggle with sub-word stutters such as sound repetition or revision, they perform well for word repetition or prolongation. This can be observed in Table \ref{table: results} as \cite{interspeech2018} performs better than our method by a small margin (3.2\%) for word repetition. Additionally, \cite{interspeech2018} performs with a lower miss rate than ours for detection of prolongation (5.92\%). Since our method relies on spectrogram features as opposed to a language model, some longer utterances can exceed the four-second windows or suffer from alignment issues, causing those stutters to be misclassified. Hence our approach produces slightly more false negatives in classification of prolongation.

As shown in Table \ref{table: results}, our classifier is able to identify interjections with an accuracy of 81.4\%. Many other works on stutter classification tend to avoid interjection disfluencies as a class, since interjection stutters tend to be more diverse and lack the consistency of repetition and prolongation stutters, making them more difficult to classify. While other works such as \cite{Mahesha2016, Mahesha2017} were able to robustly detect interjections with mel-frequency cepstral coefficients (MFCC), small subsets of the UCLASS dataset were used, preventing us from performing a fair comparison to our model.

The comparison between our proposed method and the baseline approach (using LSTM instead of bidirectional LSTM) fares similarly, with the bidirectional LSTM having slightly better or similar results for every class. This lack of significant difference between the two LSTM variations is most likely due to the fact that the feature embeddings learned using the ResNet portion of our pipeline are quite robust, accurately capturing the information required to represent different stutters. While the difference in bidirectional and unidirectional LSTMs is marginal, we opt to use the Bi-LSTM approach as the additional computational cost for Bi-LSTM is not significant. 

\begin{figure}
    \begin{center}
    \includegraphics[width=0.9\columnwidth]{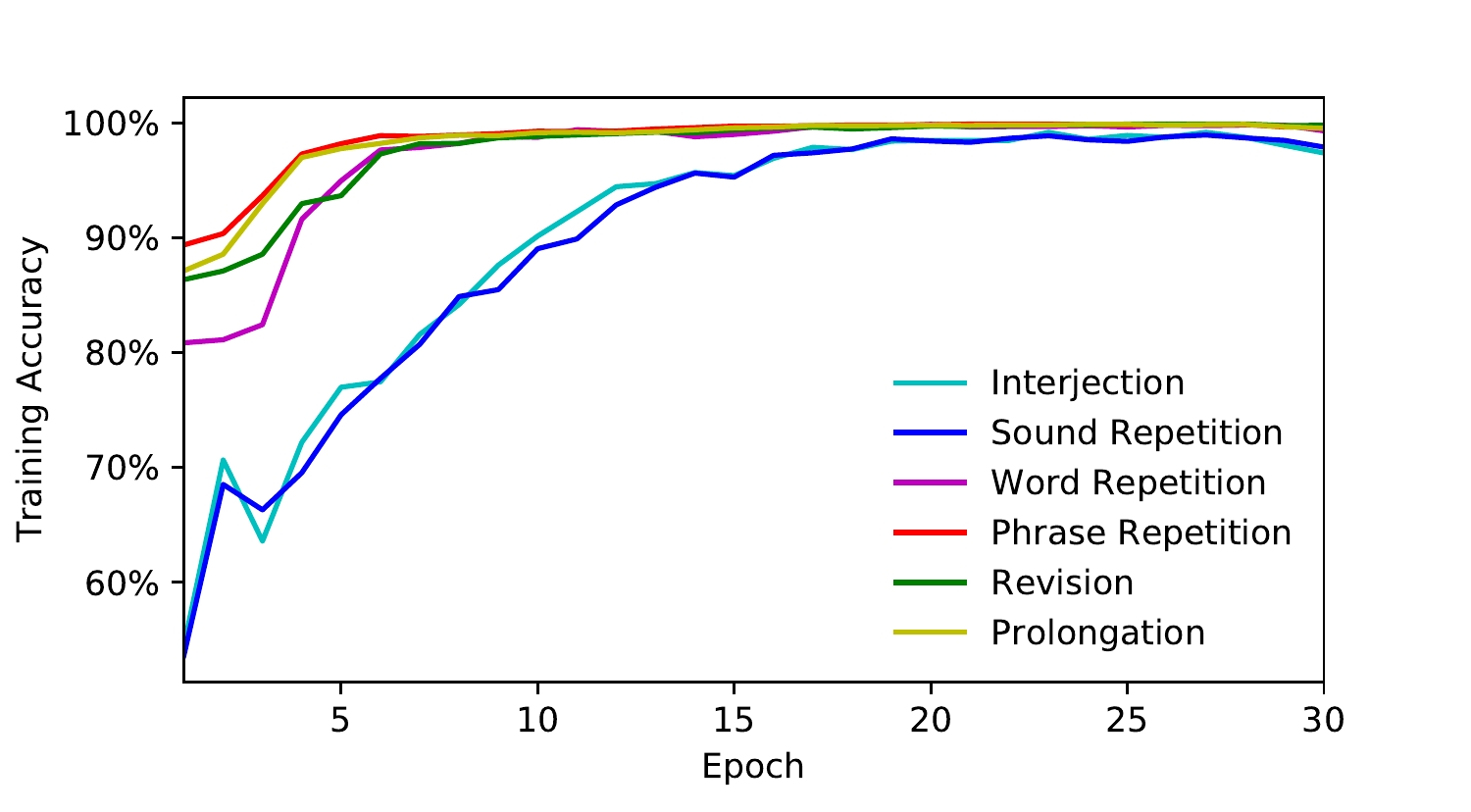}
    \end{center}
    \caption{Average training accuracy for the considered types of stutter.}
\label{fig:acc}
\end{figure}

Table \ref{table: results_ave} presents the average performance of our model compared to \cite{interspeech2018} and the baseline approach. It can be observed that our model achieves an improvement of 26.97\% lower miss rate on the UCLASS dataset over the previous state-of-the-art. Moreover, as previously shown, our method slightly outperforms the unidirectional LSTM baseline when averaged across all stutter types. Lastly, Figure \ref{fig:acc} shows the performance of our method for different stutter types against different training epochs. It can be seen that after approximately 20 epochs, our model reaches a steady-state, indicating stable learning of disfluency-related features throughout the learning phase.

\section{Conclusion and Future Work}
We present a method for detection and classification of different types of stutter disfluencies. Our model utilizes a residual network and bidirectional LSTM units trained using input spectrogram features calculated from labeled audio segments of stuttered speech. Six classes of stutter were considered in this paper: sound repetition, word repetition, phrase repetition, revision, interjection, and prolongation. Investigations show that our method performs robustly across all classes and performs with very high average accuracy and low average miss rate, achieving state-of-the-art with a significant improvement over previous the previous state-of-the-art for stutter detection. 

In future work, building upon the current model, we will conduct research on multi-class learning of different stutter disfluencies. As multiple stutter types may occur at once (e.g. \textit{'I went to uh to to uh to'}), this approach may result in more robust classification of stutters.

\section{Acknowledgements} The authors would like to thank Prof. Jim Hamilton for his support and valuable feedback throughout the course of this of this work.

\small
\bibliographystyle{IEEEbib}
\bibliography{refs}

\end{document}